\documentclass[12pt, reqno]{amsart}
\usepackage{amsmath, amsthm, fullpage}

\newtheorem{thm}{Theorem}
\newtheorem{lemma}[thm]{Lemma}
\newtheorem{theorem}[thm]{Theorem}
\theoremstyle{definition}
\newtheorem{example}[thm]{Example}

\numberwithin{equation}{section}

\newcommand{\la}{\lambda}
\renewcommand{\phi}{\varphi}

\newcommand{\R}{{\mathbb R}}
\newcommand{\Gmin}{\Gamma_{\rm{min}}}

\renewcommand{\qed}{\rule{3mm}{3mm}}
\renewenvironment{proof}
    {\vspace{1mm}\noindent\textbf{Proof.}}
    {\hspace*{\fill} $\qed$\vspace{1mm}}

\begin{document}
\title{Dispersion relation for water waves\\ with non-constant vorticity}
\author{Paschalis Karageorgis}
\address{School of Mathematics, Trinity College, Dublin 2, Ireland}
\email{pete@maths.tcd.ie}

\begin{abstract}
We derive the dispersion relation for linearized small-amplitude gravity waves for various choices of non-constant vorticity.
To the best of our knowledge, this relation is only known explicitly in the case of constant vorticity.  We provide a wide
range of examples including polynomial, exponential, trigonometric and hyperbolic vorticity functions.

\end{abstract}
\maketitle

\section{Introduction}
Consider the classical water wave problem with a free surface over a flat bottom under the influence of gravity.  Using
bifurcation and degree theory, Constantin and Strauss \cite{CS} proved the existence of large-amplitude, two-dimensional,
inviscid, periodic traveling waves with vorticity.  A relation that arises in their work, also known as the dispersion
relation, is the necessary and sufficient condition for local bifurcation.  It is known explicitly in the case of constant
vorticity \cite{CS, KMP} but not for any other vorticity functions.  Our goal in this paper is to provide a wide range of
examples of non-constant vorticity functions for which the dispersion relation can be determined explicitly.

Before we go on, however, let us first summarize the existence result of Constantin and Strauss \cite{CS}.  For the sake of
brevity, we only describe the details which are most relevant to our work and refer the reader to \cite{CS} for more details.
We let $(u,v)$ denote the velocity field and we assume that $u<c$, where $c$ is the wave speed.  This assumption makes the
relative mass flux $p_0$ negative, and it also allows us to express the vorticity $\omega=\gamma(\psi)$ in terms of the stream
function $\psi$.  As a preliminary result to local bifurcation, we first recall the existence of a curve of trivial solutions
(parallel shear flows), which are gravity waves with the bottom at $y=0$ and the top at $y=d$. Following \cite{CS}, let us
also introduce the notation
\begin{equation}\label{Gamma}
\Gamma(p) = \int_0^p \gamma(-s) \,ds, \qquad \Gmin = \min_{[p_0,0]} \Gamma (p) \leq 0.
\end{equation}
For a proof of the following two results, we refer the reader to \cite[Section 3.1]{CS}.

\begin{lemma}[Parallel shear flows] \label{triv}
For each $\la>-2\Gmin$, there exists a trivial solution
\begin{equation}\label{yp}
y(p) = \int_{p_0}^p \frac{ds}{\sqrt{\la+2\Gamma(s)}}
\end{equation}
which corresponds to a parallel shear flow with velocity field
\begin{equation}\label{vel}
u(y) = c - \sqrt{\la +2\Gamma(p(y))}, \qquad v(y)=0.
\end{equation}
Here, $p(y)$ is the inverse of $y(p)$, which is related to the stream function by $p(y)=-\psi(y)$, while $c>0$ is the speed of
propagation. The corresponding water depth is given by
\begin{equation}\label{d}
d = y(0) = \int_{p_0}^0 \frac{ds}{\sqrt{\la+2\Gamma(s)}} .
\end{equation}
\end{lemma}

\begin{theorem}[Existence of small-amplitude periodic solutions]
Let $k>0$ be a given wave number and assume that the vorticity function is sufficiently regular. Then there is a curve of
small-amplitude $\frac{2\pi}{k}$-periodic solutions if and only if the Sturm-Liouville problem
\begin{equation}\label{evp}
\beta M'' + \frac 3 2 \,\beta' M' = k^2M, \qquad M(p_0)=0, \qquad \la^{3/2}M'(0)=gM(0)
\end{equation}
has a nontrivial solution $M(p)$ for some $\la>-2\Gmin$.  Here, $\beta(p)$ is defined by
\begin{equation}\label{beta}
\beta(p) = \la + 2\Gamma(p)
\end{equation}
for each $p_0\leq p\leq 0$, while $g>0$ denotes the acceleration due to gravity.
\end{theorem}

Throughout this paper, we use primes to denote derivatives with respect to $p=-\psi$. The main difficulty in finding a necessary
and sufficient condition for the solvability of \eqref{evp} is that the solution of this problem is not generally explicit.  A
typical exception to this rule is the case of constant vorticity, in which case \eqref{evp} can be transformed into an ODE with
constant coefficients.  To the best of our knowledge, no other exceptions seem to be known.

The main goal of this paper is to provide a wide range of non-constant vorticity functions for which the eigenvalue problem
\eqref{evp} can be solved explicitly.  Our approach is inspired by the case of constant vorticity, however, we are able to
treat a much larger class of examples including polynomial, exponential, trigonometric and hyperbolic vorticity functions.

Our main result is Theorem \ref{main}, which is given in Section \ref{mare} together with its proof.  Here, we use a rather
technical change of variables, yet this is inspired by the case of constant vorticity that we briefly recall in the beginning of
the section.  Several applications of our main result are given in Section \ref{exam}, where Examples \ref{poly} through
\ref{exp} are discussed.

The dispersion relation provided by Theorem \ref{main} appears in \eqref{disg} and obviously applies for all examples, so we
shall not bother to restate it in each case.  We only do so in the cases of linear vorticity (Example \ref{lin}) and
exponential vorticity (Example \ref{exp}), as those are cases for which the dispersion relation \eqref{disg} can be simplified
quite a bit.

\section{The main result}\label{mare}

In this section, we present our main result regarding explicit solutions of
\begin{equation}\label{ode}
\beta(p) M''(p) + \frac 3 2 \,\beta'(p) M'(p) = k^2M(p),
\end{equation}
where $\beta(p)= \la + 2\Gamma(p)$ is given and $k>0$ is the wave number.  Our first step is to express this equation in terms
of the physical variables, see also \cite[Lemma 2.3]{HL}.  We note that our definitions \eqref{Gamma}-\eqref{d} imply
\begin{equation}\label{ders}
\frac{dp}{dy} = \sqrt{\la+2\Gamma(p(y))} = c-u(y), \qquad \frac{du}{dy} = -\Gamma'(p(y)) = - \gamma(-p(y))
\end{equation}
as well as $p(d)=0$ and $p(0)=p_0$.  Using the change of variables
\begin{equation}\label{phi}
\phi(y) = (c-u(y))\cdot M(p(y)),
\end{equation}
we can thus transform equation \eqref{ode} into the Rayleigh equation
\begin{equation}\label{Ray1}
\ddot{\phi}(y) - k^2\phi(y) = \frac{\ddot{u}(y)\phi(y)}{u(y)-c} ,
\end{equation}
where dots denote derivatives with respect to $y$.  We now recall \eqref{ders} and express $\ddot{u}(y)$ in terms of the
vorticity.  This allows us to write the Rayleigh equation \eqref{Ray1} as
\begin{equation}\label{Ray2}
\ddot{\phi}(y) - k^2\phi(y) = -\gamma'(-p(y)) \phi(y).
\end{equation}

If the vorticity is linear, then this equation is linear with constant coefficients, so it can certainly be solved explicitly.
Assume the vorticity is constant for simplicity.  If we impose the boundary condition $M(p_0)=0$ as in \eqref{evp}, then
$\phi(0)=0$ by \eqref{phi} and so
\begin{equation*}
\phi(y) = c_1 \sinh(ky)
\end{equation*}
for some constant $c_1$.  The corresponding solution of the original problem \eqref{ode} is
\begin{equation}\label{Mex1}
M(p) = \frac{c_1\sinh(ky(p))}{\sqrt{\la+2\Gamma(p)}} = \frac{c_1}{\sqrt{\beta(p)}} \cdot
\sinh \left( \int_{p_0}^p \frac{k\,ds}{\sqrt{\beta(s)}} \right),
\end{equation}
where $\beta(p)= \la + 2\Gamma(p)$, and every solution of \eqref{ode} with $M(p_0)=0$ has this form.

The computation above shows that the solution of the eigenvalue problem \eqref{evp} can be determined explicitly, if the
vorticity is linear.  To extend this observation to more general vorticity functions, we now look for solutions of
\eqref{evp} that have the form
\begin{equation}\label{Mex2}
M(p) = \sqrt{ \frac{G(p)}{\beta(p)} } \cdot \sinh H(p)
\end{equation}
for some functions $G,H$.  As it turns out, the unknown functions $G,H$ must be related in a very simple way.  More precisely,
\eqref{Mex2} satisfies the ODE in \eqref{evp} if and only if
\begin{align}\label{coef}
\left[ 2\beta GG'' -\beta (G')^2 + \beta' GG' - 2\beta'' G^2 + 4\beta G^2(H')^2 - 4k^2 G^2 \right] &\sinh H \notag \\
+\left[ 4\beta G^2 H'' + 4 \beta GG' H' + 2\beta' G^2 H' \right] &\cosh H = 0.
\end{align}
To ensure that the coefficient of $\cosh H$ is zero, we have to ensure that
\begin{equation*}
\frac{H''}{H'} + \frac{G'}{G} + \frac{\beta'}{2\beta} = 0
\end{equation*}
and this can be trivially integrated to yield the equivalent condition
\begin{equation}\label{HG}
H'(p) G(p) \sqrt{\beta(p)} = a
\end{equation}
for some $a\in\R$.  Assuming \eqref{HG} and returning to \eqref{coef}, we end up with the ODE
\begin{equation}\label{aode}
2\beta GG'' -\beta (G')^2 + \beta' GG' - 2\beta'' G^2 = 4k^2 G^2 -4a^2.
\end{equation}
As an ODE for $\beta$, this can be easily solved by writing the left hand side as
\begin{equation*}
2\beta GG'' -\beta (G')^2 + \beta' GG' - 2\beta'' G^2 = -2G^{3/2} \left( G^{3/2}\left( \frac{\beta}{G} \right)' \right)'
\end{equation*}
and then integrating twice.  This approach already generates a large number of examples, but one can also replace the
hyperbolic sine in \eqref{Mex2} by a regular sine, in which case \eqref{HG} is still applicable and the only change in
\eqref{aode} is that $-4a^2$ becomes $4a^2$.  We shall actually treat these cases together in our next (and main) result.

\begin{theorem}\label{main}
Let $p_0<0<k$.  Suppose $\beta(p)$, $G(p)$ are positive and $\mathcal{C}^2$ on $[p_0,0]$ with
\begin{equation}\label{odet}
\left( G^{3/2}\left( \frac{\beta}{G} \right)' \right)' = \frac{2z- 2k^2G^2}{G^{3/2}}
\end{equation}
for some constant $z\in\R$.  Define the bifurcation parameter $\la$ and the vorticity $\gamma$ by
\begin{equation}\label{laga}
\la := \beta(0), \qquad \gamma(\psi) = \gamma(-p) := \frac{1}{2}\beta'(p)
\end{equation}
and consider the eigenvalue problem \eqref{evp} which corresponds to this vorticity.  Then \eqref{evp} is solvable if and only
if the dispersion relation
\begin{equation}\label{disg}
\frac{g}{\la^{3/2}} = \frac{G'(0)}{2G(0)} - \frac{\gamma(0)}{\la} + \frac{1}{\la^{1/2} G(0) F(p_0)}
\end{equation}
holds, where $F(p_0)$ is given by
\begin{equation}\label{Fp}
F(p_0) = \left\{
\begin{array}{ccc}
\frac{1}{\sqrt z} \tanh \int_{p_0}^0 \frac{\sqrt z\,ds}{G(s)\sqrt{\beta(s)}} &&\text{if\, $z>0$} \\
\int_{p_0}^0 \frac{ds}{G(s)\sqrt{\beta(s)}} &&\text{if\, $z=0$} \\
\frac{1}{\sqrt{|z|}} \tan \int_{p_0}^0 \frac{\sqrt{|z|}\,ds}{G(s)\sqrt{\beta(s)}} &&\text{if\, $z<0$}
\end{array} \right\}.
\end{equation}
\end{theorem}
There are several things that are worth noting about this theorem.

1. Given any function $G$, equation \eqref{odet} can be easily integrated twice to obtain $\beta$, and this also determines
the vorticity $\gamma$ by means of \eqref{laga}.  Given any function $G$, in particular, one can find a vorticity function
$\gamma$ for which the dispersion relation is given by \eqref{disg}.

2. In practice, one would like to determine the function $G$ in terms of the vorticity $\gamma$ and not the other way around.
This is generally a difficult task because it involves solving a nonlinear instead of a linear ODE.  Nonetheless, it is still
feasible for a wide range of vorticity functions, as we show in Examples \ref{poly} through \ref{exp}.

3. The integral \eqref{Fp} that appears in the dispersion relation is an elementary integral only for two of our eight
examples; see the case of linear vorticity (Example \ref{lin}) and that of exponential vorticity (Example \ref{exp}).

4. Our assumption that $\beta$ is positive is equivalent to the assumption $\la>-2\Gmin$ that is needed for the trivial flows
of Lemma \ref{triv} to be defined.  Our assumption that $G$ is positive can easily be replaced by the assumption that $G$ has
one sign; namely, equation \eqref{odet} is equivalent to equation \eqref{odet2} and $G$ satisfies the latter if and only if
$-G$ does.  In fact, our assumption on $G$ is not method-driven; it is closely related to the fact that the eigenvalue problem
\eqref{evp} describes the associated ground state, see \cite{CS}.

5. The bifurcation parameter $\la$ that appears in the dispersion relation \eqref{disg} can also be expressed in terms of the
shear flows of Lemma \ref{triv}, namely $\sqrt{\la} = c-u(d)$ by \eqref{vel}-\eqref{d}.

6. The dispersion relation \eqref{disg} is the necessary and sufficient condition for the solvability of the eigenvalue
problem \eqref{evp}.  The actual solution of this problem appears in \eqref{z>0} for the case $z>0$, in \eqref{z<0} for the
case $z<0$ and in \eqref{z=0} for the case $z=0$.

\begin{proof}
We temporarily ignore one of the boundary conditions in \eqref{evp} and focus on
\begin{equation}\label{pr1}
\beta M'' + \frac 3 2 \,\beta' M' = k^2M, \qquad M(p_0)=0.
\end{equation}
To solve this explicitly, we recall that $\beta,G$ are positive and we change variables by
\begin{equation}\label{pr2}
M(p) = \sqrt{ \frac{G(p)}{\beta(p)} } \cdot \Psi(p).
\end{equation}
Then a short computation shows that \eqref{pr1} is equivalent to
\begin{align}\label{pr3}
4\beta G^2 \Psi'' &+ 2(2\beta G G' + \beta' G^2) \Psi' \notag \\
&+ \left( 2\beta GG'' - \beta (G')^2 + \beta'GG' - 2\beta'' G^2 - 4k^2 G^2 \right) \Psi = 0
\end{align}
subject to $\Psi(p_0)=0$.  Note that our assumption \eqref{odet} can also be written as
\begin{equation}\label{odet2}
2\beta GG'' - \beta (G')^2 + \beta'GG' - 2\beta'' G^2 - 4k^2G^2 + 4z =0.
\end{equation}
This allows us to reduce the problem \eqref{pr1} to the equivalent problem
\begin{equation}\label{odel}
2\beta G^2 \Psi'' + (2\beta G G' + \beta' G^2) \Psi' - 2z \Psi = 0, \qquad \Psi(p_0)=0.
\end{equation}

Suppose first that $z>0$.  Then the general solution of the last ODE is given by
\begin{equation*}
\Psi(p) = c_1 \sinh \left( \int_{p_0}^p \frac{\sqrt{z} \,ds}{G(s)\sqrt{\beta(s)}} \right) + c_2
\cosh \left( \int_{p_0}^p \frac{\sqrt{z} \,ds}{G(s)\sqrt{\beta(s)}} \right)
\end{equation*}
and the boundary condition $\Psi(p_0)=0$ requires that $c_2=0$.  In particular, every solution of the problem \eqref{pr1} has
the form
\begin{equation}\label{z>0}
M(p) = \sqrt{ \frac{G(p)}{\beta(p)} } \cdot c_1 \sinh \left( \int_{p_0}^p \frac{\sqrt{z}\,ds}{G(s)\sqrt{\beta(s)}} \right)
\end{equation}
for some constant $c_1$.  We now turn to the original eigenvalue problem \eqref{evp}, namely \eqref{pr1} with an additional
boundary condition which reads
\begin{equation*}
\frac{M'(0)}{M(0)} = \frac{g}{\la^{3/2}}.
\end{equation*}
Using logarithmic differentiation, we see that \eqref{z>0} satisfies this condition if and only if
\begin{equation*}
\frac{g}{\la^{3/2}} = \frac{G'(0)}{2G(0)} - \frac{\beta'(0)}{2\beta(0)} + \frac{1}{G(0) \sqrt{\beta(0)} F(p_0)},
\end{equation*}
where $F(p_0)$ is given by \eqref{Fp}.  Since $\beta(0)=\la$ and $\beta'(0)=2\gamma(0)$, the result follows.

When $z<0$, our previous approach applies with minor changes to give
\begin{equation}\label{z<0}
M(p) = \sqrt{ \frac{G(p)}{\beta(p)} } \cdot c_1 \sin \left( \int_{p_0}^p \frac{\sqrt{|z|} \,ds}{G(s)\sqrt{\beta(s)}} \right)
\end{equation}
instead of \eqref{z>0}, while the analogous solution for the case $z=0$ is
\begin{equation}\label{z=0}
M(p) = \sqrt{ \frac{G(p)}{\beta(p)} } \cdot c_1 \int_{p_0}^p \frac{ds}{G(s)\sqrt{\beta(s)}}.
\end{equation}
In either of these cases, the dispersion relation can be obtained exactly as before.
\end{proof}

\section{Some examples}\label{exam}
In this section, we give several applications of Theorem \ref{main}.  More precisely, we give several choices of positive
functions $\beta,G$ such that \eqref{odet} holds.  In each case, the dispersion relation is given by \eqref{disg} and we shall
only bother to restate it for Examples \ref{lin} and \ref{exp}, namely the ones for which the dispersion relation can be
simplified to a large extent.

In our first two examples, we start with a function $G$ and find a vorticity function $\gamma$ for which our theorem applies.
In all other examples, we follow the exact opposite approach and start with the vorticity function, instead.

\begin{example}[Polynomial vorticity]\label{poly}
Let $G(p)=(1-ap)^n$ for some $a>0$ and some real number $n$ other than $\pm 2,2/3$.  Integrating \eqref{odet} twice gives
\begin{equation*}
\beta(p) = 2c_1 (1-ap)^n + \frac{2c_2(1-ap)^{1-n/2}}{a(3n-2)} + \frac{4z(1-ap)^{2-2n}}{a^2(3n-2)^2} + \frac{4k^2 (1-ap)^2}{a^2(n^2-4)}.
\end{equation*}
The coefficients $c_1,c_2,z$ can be arbitrary, but we do need $\beta(p)$ to be positive on $[p_0,0]$.  An easy way to ensure
this is to assume $c_1,c_2,z$ are non-negative and $n>2$.  The vorticity $\gamma$ is then determined using \eqref{laga}. In
the special case $c_2=z=0$, one finds that
\begin{equation*}
\gamma(\psi)= -anc_1 (1+a\psi)^{n-1} - \frac{4k^2 (1+a\psi)}{a(n^2-4)},
\end{equation*}
where $\psi=-p$ as usual; the only restrictions in this case are $a>0$, $c_1\geq 0$ and $n>2$.
\end{example}

\begin{example}[Exponential vorticity \#1]
Let $G(p)= e^{ap}$ for some $a\neq 0$ and let $z\in\R$ be arbitrary.  Then \eqref{odet} can be easily integrated twice to give
\begin{equation*}
\beta(p) = c_1 e^{ap} + c_2 e^{-ap/2} + \frac{4k^2}{a^2} + \frac{4ze^{-2ap}}{9a^2}.
\end{equation*}
If we require $c_1,c_2,z$ to be non-negative, then $\beta$ is positive and our theorem applies.  Once again, the corresponding
vorticity is given by \eqref{laga}, namely
\begin{equation*}
\gamma(\psi) = \frac{ac_1}{2e^{a\psi}} - \frac{ac_2 e^{a\psi/2}}{4} - \frac{4ze^{2a\psi}}{9a}.
\end{equation*}
\end{example}

\begin{example}[Linear vorticity]\label{lin}
Suppose $\gamma(\psi) = a\psi+b$ for some $a,b\in\R$.  Then
\begin{equation*}
\beta(p) = \la + 2\int_0^p (b-as) \,ds = \lambda + 2bp - ap^2
\end{equation*}
so $\beta''(p) = -2a$.  In particular, \eqref{odet} holds with $G(p)=1$ and $z=k^2-a$.  The dispersion relation \eqref{disg}
provided by Theorem \ref{main} involves the integral
\begin{equation*}
\int_{p_0}^0 \frac{ds}{G(s)\sqrt{\beta(s)}} = \int_{p_0}^0 \frac{ds}{\sqrt{\la + 2\Gamma(s)}},
\end{equation*}
which is equal to the depth $d$ by \eqref{d}.  Thus, the dispersion relation \eqref{disg} reduces to
\begin{equation}\label{DR1}
\la - bF(d) \la^{1/2} - gF(d) = 0,
\end{equation}
where $F(d)$ is given by \eqref{Fp}, namely
\begin{equation*}
F(d) = \left\{
\begin{array}{ccc}
\frac{\tanh(d\sqrt{k^2-a})}{\sqrt{k^2-a}} &&\text{if\, $a<k^2$} \\
d &&\text{if\, $a=k^2$} \\
\frac{\tan(d\sqrt{a-k^2})}{\sqrt{a-k^2}} &&\text{if\, $a>k^2$}
\end{array} \right\}.
\end{equation*}
Needless to say, one can also solve the quadratic equation \eqref{DR1} and write it as
\begin{equation}\label{DR2}
\sqrt\la = \frac{bF(d)}{2} \pm \frac 1 2 \sqrt{b^2 F(d)^2 + 4gF(d)},
\end{equation}
where the sign is chosen so that $\sqrt{\la} = c-u(d)$ is positive.  To the best of our knowledge, this dispersion relation
was only known in the case of constant vorticity, in which case it goes back to \cite{KMP}, see also \cite{CS, Ka}.
\end{example}

\begin{example}[Quadratic vorticity]\label{quad}
Suppose $\gamma(\psi) = a\psi^2$ for some $a\neq 0$ and let
\begin{equation*}
G(p)= \frac{50a^3p^3}{3} - 20a^2 k^2 p^2 + 16ak^4 p + (25a^2\la -16k^6), \qquad \beta(p)= \la + \frac{2ap^3}{3}.
\end{equation*}
Then one can verify that \eqref{odet2}, or equivalently \eqref{odet}, holds with
\begin{equation*}
z= k^2( 1125 a^4\la^2 - 1056 a^2 k^6 \la + 256k^{12}) > 0.
\end{equation*}
We actually came up with this example by looking for solutions of \eqref{odet2} for which $\beta,G$ are both polynomials.  The
same approach applies to any quadratic vorticity whatsoever, but the corresponding cubic solution $G$ looks much messier in
the general case.  Note that $\beta$ is monotonic for this example and that the same is true for $G$, since
\begin{equation*}
G'(p) = 2a (25a^2 p^2 - 20ak^2 p + 8k^4)
\end{equation*}
and the discriminant of the quadratic is $-400 a^2 k^4$.  In particular, positivity of $\beta,G$ on the interval $[p_0,0]$
amounts to positivity of $\beta,G$ at the endpoints of the interval.
\end{example}

\begin{example}[Cubic vorticity]
Suppose $\gamma(\psi) = a\psi^3 +b\psi$ for some $a,b\in\R$ and let
\begin{equation*}
G(p)= \frac{9a^2p^4}{2} +3a(k^2 + 3b) p^2 + (2k^4 + 6bk^2 -9a\la), \qquad \beta(p)= \la - \frac{ap^4}{2} - bp^2.
\end{equation*}
Then one can verify that \eqref{odet2} holds with
\begin{equation*}
z= 2k^2 (2k^4 + 6bk^2 - 9a\la) (k^4 +2bk^2 -6a\la - 3b^2).
\end{equation*}
This example was also obtained by looking for polynomial solutions of \eqref{odet2}, an approach that no longer works for a
more general cubic or even a quartic vorticity.  To ensure that our theorem is applicable on any interval $[p_0,0]$
whatsoever, we impose the conditions
\begin{equation}\label{cond}
G(0)= 2k^2(k^2 +3b) - 9a\la > 0;\qquad a< 0 < \la; \qquad b\leq 0.
\end{equation}
Assuming those, $\beta$ is decreasing with $\beta(0)>0$, so it is positive on $[p_0,0]$, while
\begin{equation*}
G'(p) = 6ap(3ap^2 + k^2 +3b).
\end{equation*}
If it happens that $k^2+3b\leq 0$, then $G$ is also decreasing with $G(0)>0$, so we are done.  If it happens that $k^2+3b>0$,
then $G$ has a unique critical point $p_*<0$, where
\begin{equation*}
p_*= -\sqrt\frac{k^2 +3b}{3|a|}, \qquad G(p_*) = -9a\la +\frac{3}{2}\,(k^2+3b)(k^2-b) > 0.
\end{equation*}
In either case then, $G$ is positive on the whole negative real axis, as needed.  To avoid the technical condition in
\eqref{cond}, one may focus on the special case $a<0<\la$ and $b=0$.
\end{example}

\begin{example}[Trigonometric vorticity]
Suppose $\gamma(\psi)= a_1\cos(b\psi) +a_2\sin(b\psi)$ for some $a_1,a_2\in\R$ and some $b>0$.  In this case, one can take
\begin{equation*}
G(p)= \beta(p) + \frac{4k^2}{b^2}, \qquad \beta(p) = \lambda + \frac{2a_1\sin(bp)}{b} + \frac{2a_2(\cos(bp)-1)}{b}
\end{equation*}
and check that \eqref{odet2} holds with
\begin{equation*}
z= \frac{k^2}{b^4} \cdot \Bigl[ (4k^2+ \la b^2)^2 - 4a_2 b (4k^2+ \la b^2) - 4a_1^2 b^2 \Bigr].
\end{equation*}
To ensure that $\beta(p)$ is positive for all $p$, we have to impose the condition
\begin{equation*}
\la > \frac{2}{b} \left( a_2 + \sqrt{a_1^2 + a_2^2} \right),
\end{equation*}
while the positivity of $G(p)$ follows trivially from that of $\beta(p)$.  This example is related to the previous two in the
following sense.  When $a_1=0$, the eigenvalue problem \eqref{evp} can be solved using the substitution $u=\cos(bp)$ that reduces
the problem to one with polynomial coefficients; the solution $G(p)$ was found by looking for polynomial solutions in $u$.
\end{example}

\begin{example}[Hyperbolic vorticity]
Suppose $\gamma(\psi)= a_1\cosh(b\psi) +a_2\sinh(b\psi)$ for some $a_1,a_2\in\R$ and some $b>0$.  In this case, one can take
\begin{equation}\label{G1}
G(p)= \beta(p)-\frac{4k^2}{b^2}, \qquad \beta(p) = \lambda + \frac{2a_1\sinh(bp)}{b} + \frac{2a_2(1-\cosh(bp))}{b}
\end{equation}
and check that \eqref{odet2} holds with
\begin{equation}\label{h1}
z= \frac{k^2}{b^4} \cdot \Bigl[ (4k^2- \la b^2)^2 - 4a_2 b (4k^2- \la b^2) + 4a_1^2 b^2 \Bigr].
\end{equation}
Needless to say, one can always obtain this example from the previous one by allowing the coefficients to be complex.  If we
assume $a_1\leq \min \{-a_2,0\}$, then we have
\begin{equation*}
\beta'(p) = (a_1 - a_2) e^{bp} + (a_1 + a_2) e^{-bp} \leq 0
\end{equation*}
for all $p\leq 0$.  This is clear when $a_1-a_2\leq 0$, in which case $\beta'$ is the sum of two non-positive terms, but it
also holds when $a_1-a_2>0$, in which case
\begin{equation*}
\beta'(p) \leq (a_1 - a_2) e^{-bp} + (a_1 + a_2) e^{-bp} = 2a_1 e^{-bp} \leq 0.
\end{equation*}
It easily follows that our theorem applies when $a_1\leq \min \{-a_2,0\}$, $b>0$ and $\la>\frac{4k^2}{b^2}$.
\end{example}

\begin{example}[Exponential vorticity \#2]\label{exp}
Suppose $\gamma(\psi)= ae^{b\psi}$ for some $a,b\neq 0$.  This is a special case of the previous example and one can take
\begin{equation}\label{Gh}
G(p)= \beta(p)-\frac{4k^2}{b^2}, \qquad \beta(p) = \lambda - \frac{2a(e^{-bp}-1)}{b}, \qquad z=
\frac{k^2(4k^2- \la b^2 - 2ab)^2}{b^4}.
\end{equation}
In this case, the dispersion relation \eqref{disg} can be simplified quite a bit.  Suppose $z>0$ for the moment. Then
\eqref{z>0} shows that every solution of \eqref{evp} is a scalar multiple of
\begin{equation*}
M_0(p) = \sqrt{ \frac{G(p)}{\beta(p)} } \cdot \sinh \left( \int_{p_0}^p \frac{\sqrt{z} \,ds}{G(s)\sqrt{\beta(s)}} \right),
\end{equation*}
hence also a scalar multiple of
\begin{equation}\label{M1}
M_1(p) = \sqrt{ \frac{G(p)}{\beta(p)} } \cdot \sinh \left( \int_{p_0}^p \frac{k(4k^2- \la b^2 -2ab)}{b^2 G(s) \sqrt{\beta(s)}}
\:ds \right).
\end{equation}
We now recall our choice \eqref{Gh} of $G(p)$ and we make use of the identity
\begin{equation}\label{ide}
\left[ \frac{1}{2} \,\log \frac{\sqrt{\beta(p)} + 2k/b}{\sqrt{\beta(p)} -2k/b} \right]' - \frac{k}{\sqrt{\beta(p)}} =
\frac{k(4k^2- \la b^2 -2ab)}{b^2 G(p) \sqrt{\beta(p)}}
\end{equation}
that may be viewed as a partial fractions decomposition for the right hand side.  Using this identity and our choice of $G(p)$ in
\eqref{Gh}, one finds that
\begin{align}\label{M2}
M_1(p) &= \left( \frac{1}{2} +\frac{k}{b\sqrt{\beta(p)}} \right) \sqrt{\frac{\sqrt{\beta(p_0)}-2k/b}{\sqrt{\beta(p_0)}+2k/b}}
\cdot \exp \left( -\int_{p_0}^p \frac{k\,ds}{\sqrt{\beta(s)}} \right) \notag \\
&\quad - \left( \frac{1}{2} - \frac{k}{b\sqrt{\beta(p)}} \right) \sqrt{ \frac{\sqrt{\beta(p_0)}+2k/b}{\sqrt{\beta(p_0)}-2k/b}}
\cdot \exp \left( \int_{p_0}^p \frac{k\,ds}{\sqrt{\beta(s)}} \right).
\end{align}
Define the coefficients $c_1,c_2$ by the formula
\begin{equation*}
c_1 =\sqrt{\beta(p_0)} -2k/b, \qquad c_2 =\sqrt{\beta(p_0)} +2k/b.
\end{equation*}
In view of \eqref{vel}, these can also be expressed in terms of the velocity field as
\begin{equation}\label{c1c2}
c_1 = c-u(0) -2k/b, \qquad c_2 = c-u(0) +2k/b.
\end{equation}
Now, every solution of \eqref{evp} is a scalar multiple of \eqref{M2}, hence also a scalar multiple of
\begin{align}\label{M3}
M(p)= c_1 &\left( 1 + \frac{2k}{b\sqrt{\beta(p)}} \right) \cdot \exp
\left(-\int_{p_0}^p \frac{k\,ds}{\sqrt{\beta(s)}} \right) \notag \\
&\qquad -c_2 \left( 1 - \frac{2k}{b\sqrt{\beta(p)}} \right) \cdot
\exp \left(\int_{p_0}^p \frac{k\,ds}{\sqrt{\beta(s)}} \right).
\end{align}

We remark that the last formula was obtained under the assumption that $z,\beta,G$ are all positive, yet this formula provides a
solution of \eqref{evp} as long as merely $\beta$ is positive.  Next, we turn to the boundary condition at $p=0$, which reads
\begin{equation}\label{boco}
\la^{3/2} M'(0) = g M(0).
\end{equation}
Differentiating \eqref{M3}, one finds that
\begin{align*}
M'(p)= &-\frac{kc_1}{\sqrt{\beta(p)}} \left( 1 + \frac{2k}{b\sqrt{\beta(p)}} + \frac{\beta'(p)}{b\beta(p)} \right) \cdot
\exp \left(-\int_{p_0}^p \frac{k\,ds}{\sqrt{\beta(s)}} \right) \notag \\
&- \frac{kc_2}{\sqrt{\beta(p)}} \left( 1 - \frac{2k}{b\sqrt{\beta(p)}} + \frac{\beta'(p)}{b\beta(p)} \right) \cdot
\exp \left(\int_{p_0}^p \frac{k\,ds}{\sqrt{\beta(s)}} \right).
\end{align*}
In view of \eqref{Gh} and \eqref{d}, this implies
\begin{align*}
M'(0)= -\frac{kc_1 e^{-kd}}{\sqrt\la} \left( 1 + \frac{2k}{b\sqrt\la} + \frac{2a}{b\la} \right) -
\frac{kc_2 e^{kd}}{\sqrt\la} \left( 1 - \frac{2k}{b\sqrt\la} + \frac{2a}{b\la} \right)
\end{align*}
and also
\begin{equation*}
M(0)= c_1e^{-kd} \left( 1 + \frac{2k}{b\sqrt\la} \right) - c_2e^{kd} \left( 1 - \frac{2k}{b\sqrt\la} \right).
\end{equation*}
Thus, the boundary condition \eqref{boco} holds if and only if
\begin{align*}
c_1e^{-kd} \left[ (g+k\la) \left( 1 + \frac{2k}{b\sqrt\la} \right) +\frac{2ak}{b} \right] = c_2e^{kd}
\left[ (g-k\la) \left( 1 -\frac{2k}{b\sqrt\la} \right) - \frac{2ak}{b} \right].
\end{align*}
Multiplying through by $b\sqrt\la$, one thus obtains the dispersion relation
\begin{equation}\label{disexp}
c_1 \Bigl( (g+k\la)(b\sqrt\la +2k) + 2ak\sqrt\la \Bigr) = c_2e^{2kd}
\Bigl( (g-k\la)(b\sqrt\la -2k) -2ak\sqrt\la \Bigr),
\end{equation}
where $d$ is the depth, $\sqrt{\la}=c-u(d)$ and the coefficients $c_1,c_2$ are given by \eqref{c1c2}.
\end{example}

\bibliographystyle{siam}
\bibliography{dis_rel}
\end{document}